\documentclass[twocolumn]{aastex631}

\usepackage{amsmath}
\usepackage{amssymb}
\usepackage{mathrsfs}

\usepackage[T1]{fontenc}

\renewcommand{\edit}[2]{#2}

\begin{document}
\title{
A disk instability model for the quasi-periodic eruptions of GSN 069  
}


\author[0000-0002-6938-3594]{Xin Pan}
\affiliation{Key Laboratory for Research in Galaxies and Cosmology, Shanghai Astronomical Observatory, Chinese Academy of Sciences, 80 Nandan Road, Shanghai 200030, People's Republic of China}
\affiliation{University of Chinese Academy of Sciences, 19A Yuquan Road, 100049, Beijing, People's Republic of China}

\author[0000-0002-7299-4513]{Shuang-Liang Li}
\affiliation{Key Laboratory for Research in Galaxies and Cosmology, Shanghai Astronomical Observatory, Chinese Academy of Sciences, 80 Nandan Road, Shanghai 200030, People's Republic of China}
\affiliation{University of Chinese Academy of Sciences, 19A Yuquan Road, 100049, Beijing, People's Republic of China}

\author[0000-0002-2355-3498]{Xinwu Cao}
\affiliation{Zhejiang Institute of Modern Physics, Department of Physics, Zhejiang University, 38 Zheda Road, Hangzhou 310027, People's Republic of China}
\affiliation{Shanghai Astronomical Observatory, Chinese Academy of Sciences, 80 Nandan Road,  Shanghai, 200030, People's Republic of China}

\author[0000-0003-0707-4531]{Giovanni Miniutti}
\affiliation{Centro de Astrobiolog\'{i}a (CSIC-INTA), Camino Bajo del Castillo s/n, Villanueva de la Ca\~{n}ada, E-28692 Madrid, Spain}

\author[0000-0002-4455-6946]{Minfeng Gu}
\affiliation{Key Laboratory for Research in Galaxies and Cosmology, Shanghai Astronomical Observatory, Chinese Academy of Sciences, 80 Nandan Road, Shanghai 200030, People's Republic of China}

\correspondingauthor{Xin Pan, Shuang-Liang Li, Xinwu Cao}
\email{panxin@shao.ac.cn, lisl@shao.ac.cn, xwcao@zju.edu.cn}

\begin{abstract}
GSN 069 is a recently discovered QPE (Quasi-periodic eruptions) source recurring about every 9 hours. The mechanism for the QPEs of GSN 069 is still unclear so far. In this work, a disk instability model is constructed to explain GSN 069 based on \citet{2021ApJ...910...97P} (PLC21), where the authors proposed a toy model for the repeating changing-look (CL) active galactic nuclei (AGN). We improve the work of PLC21 by including a non-zero viscous torque condition on the inner boundary of disk and adopting a general form for the viscous stress torque in Kerr metric. It is found that the 0.4-2 keV light curves, the light curves at different energy bands and the phase-resolved X-ray spectrum of GSN 069 can all be qualitatively reproduced by our model. Furthermore, the profiles of light curve in QPEs can be significantly changed by the parameter $\mu$ in viscous torque equation, which implies that our model may also be applied to other QPEs.


\end{abstract}


\section{Introduction}
Quasi-periodic eruptions (QPEs) are new phenomena exhibiting quasi-periodic rapid and high-amplitude bursts in soft X-ray, which were reported by \citet{2019Natur.573..381M} firstly in a low-mass Seyfert 2 galaxy GSN 069. The burst duration and period of GSN 069 are about 1 and 9 hours, respectively. Except for GSN 069, four other QPE sources, i.e., RX J1301.9+2747 \citep{2013ApJ...768..167S, 2020A&A...636L...2G}, 2MASS 02314715-1020112, 2MASX J02344872-4419325 \citep{Arcodia_2021} and XMMSL1 J024916.6-041244 \citep{2021ApJ...921L..40C}, have been discovered recently. Compared with others, more physical properties of GSN 069 have been inferred, such as the central black hole mass ($M_{\rm BH}\sim4 \times 10^5\rm M_{\odot}$ with a factor of few uncertainty), the X-ray spectrum in various stages during outbursts and the nearly constant eruption period with alternating long/short QPE time-separation as well as alternating strong/weak QPEs. All of these are very important for us to study the physical origin of QPEs.

Several models have been proposed to explain the physics of QPEs: radiation pressure instabilities in the transition zone between inner advection-dominated accretion flow (ADAF) and outer thin disk (\citealt{2020A&A...641A.167S,2021ApJ...910...97P}, hereinafter PLC21); Roche lobe overflow from one/two stars orbiting the central black hole \citealt{2021arXiv210903471Z,2021arXiv210713015M}; mass overflow at pericentre from a white dwarf around a super-massive black hole (a near miss tidal disruption, \citealt{2020MNRAS.493L.120K}); warped disk tearing \citep{2021ApJ...909...82R}; self-lensing of binary massive black hole \citep{2021MNRAS.503.1703I}; and star-disk collisions \citep{2021ApJ...917...43S,2021ApJ...921L..32X}. These models can partly explain the outbursts of QPEs, but mainly focused on their short timescale and regular period. So far, there is no model can fit both the outburst period and the X-ray spectrum of GSN 069 simultaneously. The observed X-ray spectra at various burst stages in GSN 069 have been successfully reproduced by using a constant disk blackbody plus a variable blackbody component \citep{2019Natur.573..381M}. Together with the quasi-periodic outbursts, we notice that all these characteristics may be explained by the model of \citet{2020A&A...641A.167S} and PLC21.

The radiation pressure dominated inner region of a standard thin disk is both thermally and viscously unstable \citep{shakura_black_1973,1976MNRAS.175..613S}. Adopting some special values, this unstable region can be shrunken to a narrow zone between inner ADAF and outer thin disk \citep{2020A&A...641A.167S}. This model has been successfully applied to the repeating changing-look (CL) active galactic nuclei (AGN). Due to the important role of magnetic field playing on the formation of jet/winds, PLC21 investigated how the large-scale magnetic field affects the outbursts of unstable region and found that both the period and amplitude of outbursts can be significantly changed. However, the mass accretion rate of GSN 069 should be very high due to its high luminosity ($\sim 10^{42} \rm{erg s^{-1}}$) and a relatively small black hole mass ($\sim4 \times 10^5\rm M_{\odot}$, \citealt{2019Natur.573..381M}). In this case, the inner ADAF will disappear and the thin disk can extend to the innermost stable circular orbit (ISCO, \citealt{1998tbha.conf..148N}). The high mass accretion rate will also result on a larger unstable inner region in a thin disk and a much longer outburst period, which is inconsistent with the behaviour of QPEs. However, as shown in PLC21, this point can be resolved if we consider the effects of large-scale magnetic field, which can take away both the energy and angular momentum from disk and decrease its temperature significantly \citep{2013ApJ...765..149Cao,2014ApJ...786....6L,2019ApJ...872..149L}. In this work, we construct a disk instability model base on PLC21 to explain the quasi-periodic eruptions of GSN 069.

\section{Model}


\subsection{steady outer disk}

\begin{figure}[htbp]
   \centering
   \includegraphics[width=0.47\textwidth]{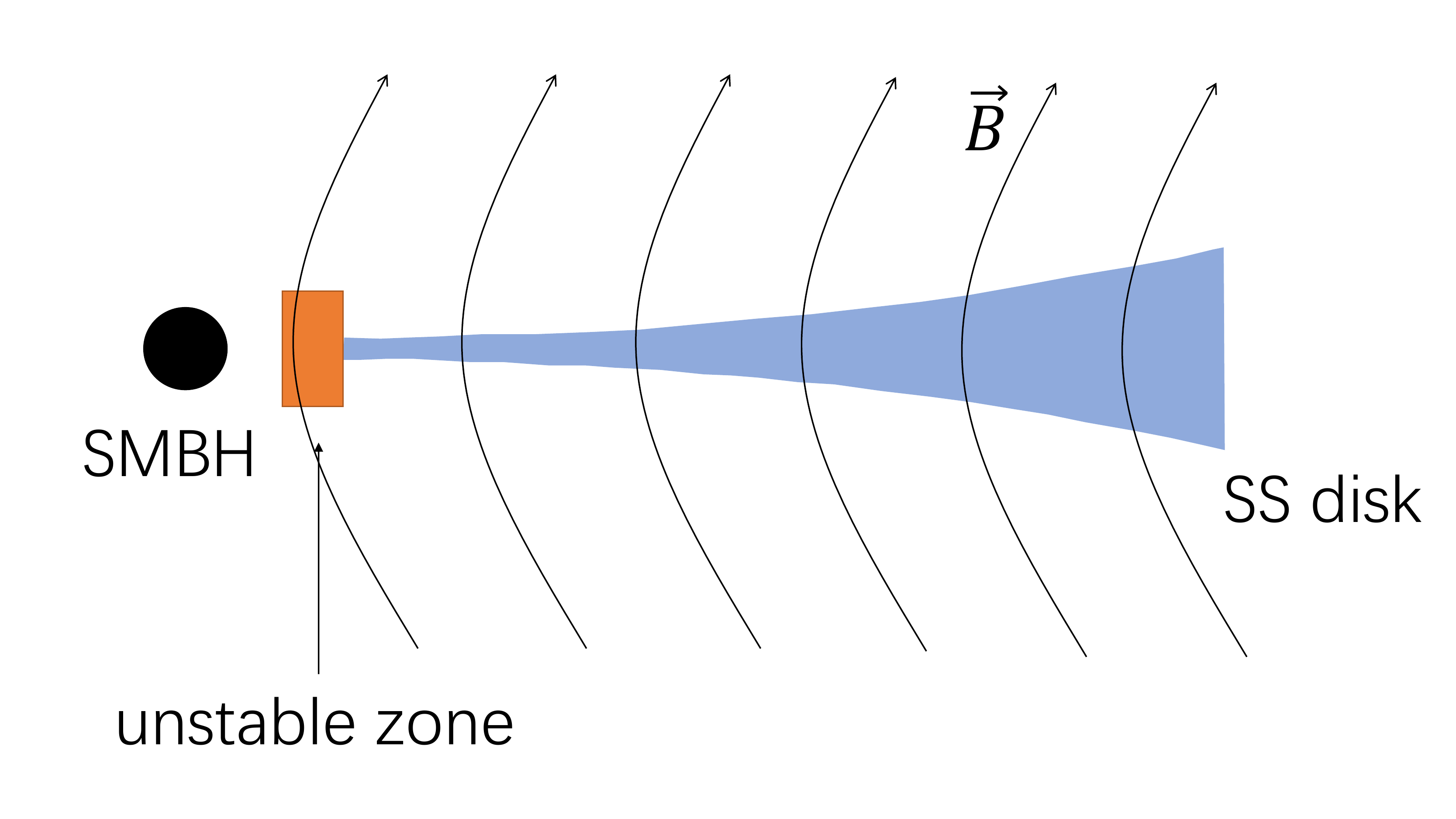}
   \caption{The schematic picture of our model. The outer blue region represents a stable thin disk dominated by gas pressure and the inner orange region is the unstable zone dominated by radiation pressure.}
   \label{fig:sch}
\end{figure}

The inner ADAF will disappear when the mass accretion rate is high enough. Therefore, the disk contains only two regions: the inner unstable and outer stable regions (see Figure \ref{fig:sch}). On the presence of large-scale magnetic field, the unstable region can be constrained to a narrow zone just outside the ISCO. In order to improve the precision of our model, we adopt the general relativistic correction factors defined by \citet{1973blho.conf..343N} to modify our equations:

\begin{equation}
    \mathscr{A}=1+\frac{a_{*}^{2}}{4\hat{r}^{2}}+\frac{a_{*}^{2}}{4\hat{r}^{3}},
    \label{A_factor}
\end{equation}

\begin{equation}
    \mathscr{B}=1+a_{*}\sqrt{\frac{1}{8\hat{r}^{3}}},
    \label{B_factor}
\end{equation}

\begin{equation}
    \mathscr{C}=1-\frac{3}{2\hat{r}}+2a_{*}\sqrt{\frac{1}{8\hat{r}^{3}}},
    \label{C_factor}
\end{equation}

\begin{equation}
    \mathscr{D}=1-\frac{1}{\hat{r}}+\frac{a_{*}^{2}}{4\hat{r}^{2}},
    \label{D_factor}
\end{equation}

\begin{equation}
    \mathscr{E}=1+\frac{a_{*}^{2}}{\hat{r}^{2}}-\frac{a_{*}^{2}}{2\hat{r}^{3}}+\frac{3a_{*}^{4}}{16\hat{r}^{4}},
    \label{E_factor}
\end{equation}

\begin{equation}
    \mathscr{F}=1-2a_{*}\sqrt{\frac{1}{8\hat{r}^{3}}}+\frac{a_{*}^{2}}{4\hat{r}^{2}},
    \label{F_factor}
\end{equation}

\begin{equation}
    \Omega_{\rm k}=\sqrt{\frac{GM}{R^3}}\frac{1}{\mathscr{B}},
    \label{Omega}
\end{equation}

\begin{equation}
    l_{\rm k}=\sqrt{GMR}\frac{\mathscr{F}}{\mathscr{C}^{1/2}},
    \label{specific_angular_momentum}
\end{equation}
where $a_{*}=cJ/GM^2$ is the dimensionless spin parameter of black hole, $\hat{r}(=R/R_{\rm s})$, $J$, $R_{\rm s}=2GM/c^2$ and $M$ are the dimensionless radius, angular momentum of black hole, Schwarzschild radius and black hole mass, respectively.

As PLC21, the general relativistic continuity equation for the outer thin disk with winds driven by large-scale magnetic field can be written by:

\begin{equation}
    \frac{\mathrm{d}\dot{M}}{\mathrm{d}R}+4\pi R\dot{m}_{\rm w}=0,
    \label{continuity}
\end{equation}
where $\dot{m}_{\rm w}$ is the mass-loss rate per unit surface area of disk (see PLC21 for the details).

The angular momentum equation can be rewritten as

\begin{equation}
    -\frac{1}{2\pi}\frac{\mathrm{d}(\dot{M}l_{\rm k})}{\mathrm{d}R}-\frac{\mathrm{d}}{\mathrm{d}R}(R^{2}\mathscr{B}\mathscr{C}^{-1/2}\mathscr{D}T_{r\phi})+T_{\rm m} R=0,
    \label{angularmom}
\end{equation}
where $T_{r\phi}$ and $T_{\rm m}$ are the viscous and the magnetic torques exerted on the accretion disk, respectively (see PLC21 for the details).

While the viscosity in accretion disk is believed to be related to the turbulence induced by the magnetorotational instability (MRI, \citealt{1998RvMP...70....1B}), the viscous torque has been suggested to be either proportional to the total pressure \citep{shakura_black_1973}, or proportional to the gas pressure \citep{1981ApJ...247...19S}, or both \citep{2009ApJ...698..840C}. In this work, we adopt a general form for the viscous torque:

\begin{equation}
    T_{r\phi}=-2\alpha P_{\rm tot}^{1-\mu}P_{\rm gas}^{\mu}H,
    \label{viscous_stress}
\end{equation}
where $P_{\rm tot}$ and $P_{\rm gas}$ are the total and gas pressure, respectively \citep{1990PASJ...42..661H,2008ApJ...683..389D}. This choice is directly motivated by the requirement that the outer radius of unstable zone should be close enough to ISCO with the parameters presented in Section 3, in order to get a narrow unstable zone. The widely used $\alpha$-viscosity is $\nu=-\alpha C_{\rm s}H(\mathrm{d ln}\Omega_{\rm k}/\mathrm{d ln}R)^{-1}$ (and $T_{r\phi}=\nu\Sigma R\mathrm{d}\Omega_{\rm k}/\mathrm{d}R$), where $C_{s}$ is sound speed of local disk and $H=(P_{\rm tot}/\rho)^{1/2}\Omega_{\rm k}^{-1}$ is the half thickness of disk. With $C_{\rm s}=(P_{\rm tot}/\rho)^{1/2}$, we have $\nu\propto\alpha P_{\rm tot}$ \citep{shakura_black_1973}, corresponding to $\mu=0$ in equation \ref{viscous_stress}. \citet{1984ApJ...287..761T} proposed that the turbulent velocity may be related to sound speed $C_{s}=(P_{\rm gas}/\rho)^{1/2}$ instead of $C_{s}=(P_{\rm tot}/\rho)^{1/2}$, which leads to $\nu\propto\alpha(P_{\rm gas}P_{\rm tot})^{1/2}$, i.e., $\mu=0.5$. In this case, the thermal instability of the disk is mostly suppressed. The parameter of $\mu$ in this work is employed to describe to what extent the turbulent velocity depends on the gas pressure. The exact value of $\mu$ is currently unknown, as is its physical interpretation. Stability criteria applied to X-ray binary systems suggest $\mu\sim0.56$ (see \citet{2008ApJ...683..389D} for details). However, in this work we are not attempting to model the general behavior of accreting black holes, but rather to seek a possible explanation of a relatively rare phenomenon (QPEs) in terms of disk instabilities. As such, as discussed in Section 3, we adopt a value of $\mu=0.27$ that allows for a small instability region close to the ISCO without affecting the stability of the outer disk.

The total pressure reads:
\begin{equation}
    P_{\rm tot}=(1+\frac{1}{\beta_1})(P_{\rm gas}+P_{\rm rad}),
    \label{EoS}
\end{equation}
where $\beta_{1}=(P_{\rm gas} + P_{\rm rad})/(B^2/8\pi)$, $B$ and $P_{\rm rad}$ are magnetic strength and radiation pressure, respectively. In the standard thin disk model, the zero viscous torque condition is usually adopted at ISCO. However, an additional magnetic torque would be exerted on the ISCO when large scale magnetic fields are present (e.g., \citealt{1999ApJ...515L..73K}). The non-zero torque on the inner boundary will increase the radiative efficiency by producing additional dissipation (e.g., \citealt{2000ApJ...528..161A}). Following \citet{2000ApJ...528..161A}, we define a parameter $f$ in our model to describe the effect of non-zero torque condition at ISCO in Kerr metric:

\begin{equation}
    T_{r\phi}\bigg|_{R=R_{\rm ISCO}}=-(1-f)\left[\frac{\dot{M}l_{\rm k}\mathscr{C}^{1/2}}{2\pi\mathscr{BD}R^{2}}\right]_{R=R_{\rm ISCO}},
    \label{non_zero_torque}
\end{equation}
where the parameter $f$ can be related to the parameter $f_{\rm ms}$ in \citet{2000ApJ...528..161A} as:
\begin{equation}
f_{\rm ms}=\frac{3(1-f)}{2x_{\rm ms}^2\mathscr{C}^{1/2}_{\rm ms}}.
\end{equation}
 Our $f$ parameter can take any value between $-\infty$ and 1, while $f_{ms}$ is limited to vary between 0 and $\infty$. The zero-torque condition corresponds to $f=1$ in our model instead of $f_{ms}=0$.

Without an inner ADAF region, the energy equation will return to the classical form as:

\begin{equation}
    -\frac{3}{2}\Omega_{\rm k}T_{r\phi}\frac{\mathscr{BD}}{\mathscr{C}}=\frac{8acT_{\rm c}^{4}}{3\tau}.
    \label{energy}
\end{equation}
The optical depth $\tau$ is given by $\tau=\Bar{\kappa}\Sigma/2$, where $\Bar{\kappa}$ and $\Sigma=2\rho H$ are the opacity and surface density, respectively.



\subsection{variable inner unstable region}

The instability of a standard thin accretion disk in Kerr metric had been investigated by \citet{2011ApJS..195....7X}, where the continuity equation is given as :

\begin{equation} 
u^t \frac{\partial \Sigma}{\partial t}=-\frac{1}{r} \frac{\partial}{\partial r}(\Sigma r u^r)-\Sigma \frac{\partial u^t}{\partial t},
\label{sur_den_evolution}
\end{equation}
where $u^{t}$ and $u^{r}$ are the time and radial components of four-velocity, respectively. Similar with PLC21, the equation (\ref{sur_den_evolution}) can be rewritten as:

\begin{equation}
\begin{split}
    &\left[u^t-\frac{C_{\rm H}H\left(1-\beta_{2}\right)}{\Sigma\left(1+\beta_{2}\right)}\right]\frac{\mathrm{d}\Sigma}{\mathrm{d}t}+\frac{C_{\rm H}H\left(4-3\beta_{2}\right)}{T\left(1+\beta_{2}\right)}\frac{\mathrm{d}T}{\mathrm{d}t}\\
    &-\frac{\dot{M}_{0}-\dot{M}-4\pi R\dot{m}_{\rm w}\Delta R}{2\pi R\Delta R}=0,
    \label{sur_den_evolution_2}
\end{split}
\end{equation}
where $\dot{M}_{0}$ and $\Delta R$ are the inflow rate and the width of the unstable zone, respectively. $\beta_{2}=P_{\rm gas}/(P_{\rm gas}+P_{\rm rad})$ is the ratio of gas pressure to the sum of gas pressure and radiation pressure. In order to simplify the form of equation, we define a coefficient $C_{\rm H}$ as:

\begin{equation}
    C_{\rm H}=-\frac{u^{t}u^{r}\Sigma\dot{M}_{0}}{2\pi RH_{0}\Sigma_{0}\left[c^2\mathscr{D}+(u^r)^2\right]},
    \label{C_H}
\end{equation}
where $H_{\rm 0}$ and $\Sigma_{\rm 0}$ are the scale-height and surface density on the outer boundary of unstable zone, respectively.

The energy conservation equation is \citep{2011ApJS..195....7X}:


\begin{equation}
    \Sigma T\left[u^{t}\frac{\partial S}{\partial t}+u^{r}\frac{\partial S}{\partial r}\right]=Q^{+}-Q^{-},
    \label{temp_evolution}
\end{equation}
where $S$, $Q^{+}$ and $Q^{-}$ are the entropy of gas, the viscous heating rate and the cooling rate, respectively. The second term in the left side of Equation (\ref{temp_evolution}) represents the advection rate of energy $Q_{\rm adv}$, which can be written as (see PLC21):

\begin{equation}
    Q_{\rm adv}=\frac{\dot{M}PH}{2\pi R\Delta R\Sigma\mathscr{D}^{\frac{1}{2}}}.
    \label{Q_adv}
\end{equation}
Therefore, the evolution equation of temperature can be rewritten as:

\begin{equation}
\begin{aligned}
    \frac{\mathrm{d} T}{\mathrm{d}t}=&\frac{T(Q^+-Q^--Q_{\rm adv})(1+\frac{1}{\beta_1})(1+\beta_2)}{2PHu^{t}(28-22.5\beta_2-1.5\beta_2^2+\frac{12-9\beta_2}{\beta_1})}\\
    &+2\frac{T\mathrm{d}\Sigma}{\Sigma\mathrm{d}t}\frac{4-3\beta_2+\frac{2-\beta_2}{\beta_1}}{28-22.5\beta_2-1.5\beta_2^2+\frac{12-9\beta_2}{\beta_1}}.
    \label{temp_evolution_2}
\end{aligned}
\end{equation}

\section{Numerical Results}

With equations (\ref{sur_den_evolution_2}) and (\ref{temp_evolution_2}), we can study the limit-cycle behaviour of thin disk. The overall model is complex and depends on several free parameters. Exploring the full parameter space to compare the model with the available X-ray data would require to generate a very large number of simulations that would take decades to run. Moreover, the large number of free parameters that is inherent to any disk instability model implies that general model predictions can be obtained, but that different details can be produced easily. This  suggests to look for qualitative comparisons with the data rather than detailed fits, at least in this initial stage. We adopt here a fiducial model with a black hole mass of $2\times 10^{5}\rm M_{\odot}$ and a high black hole spin value of 0.98. A discussion on the choice of parameters for the fiducial model is given in Section 4. This value of mass is 2 times smaller than that given by \citet{2019Natur.573..381M}, but is still range within the estimated black hole mass. Other parameters are conveniently chosen as $f=0.9$, $\dot{m}=0.08\dot{M}_{\rm Edd}$ ($\dot{M}_{\rm Edd}=1.5 \times 10^{18} M/M_{\odot} \rm{ g s^{-1}}$), $\alpha=0.15$, $\beta_{1}=38$, $\mu=0.27$  and $\Delta R=0.12 R_{\rm s}$ (see Group 1 in Table 1). Here the outer radius of $\Delta R$ is given self-consistently by the disk instability criteria (see section 4 for details).

\begin{figure}[htbp]
   \centering
   \includegraphics[width=0.47\textwidth]{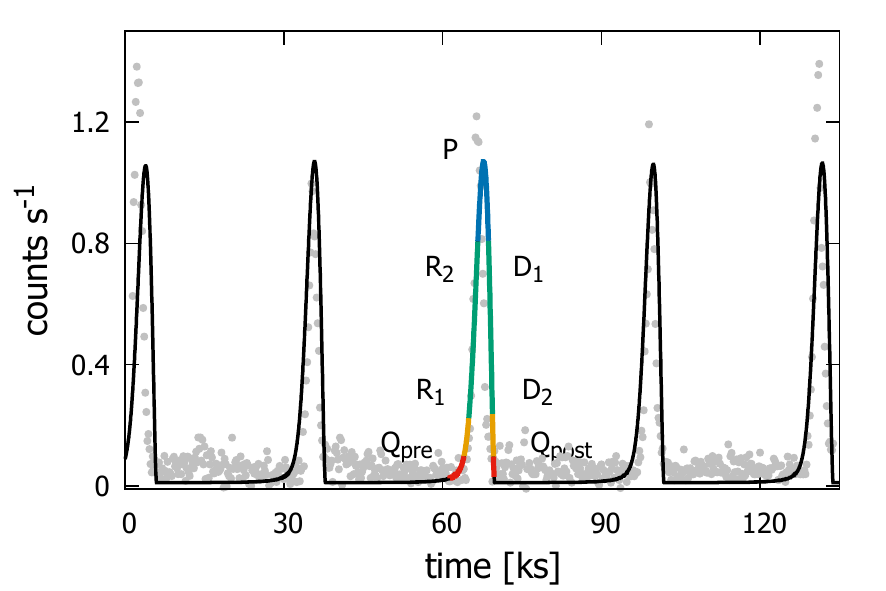}
   \caption{The 0.4-2 keV light curve, where the black line represents the light curve given by our model and the colored lines are the seven phase segments of eruption to fit the X-ray spectrum in Figure \ref{fig:spectrum}. The grey dots are the data observed by XMM-Newton Director Discretionary Time (DDT) performed on 2019 January 16/17.}
   \label{fig:light_curve}
\end{figure}

\begin{figure}[htbp]
   \centering
   \includegraphics[width=0.47\textwidth]{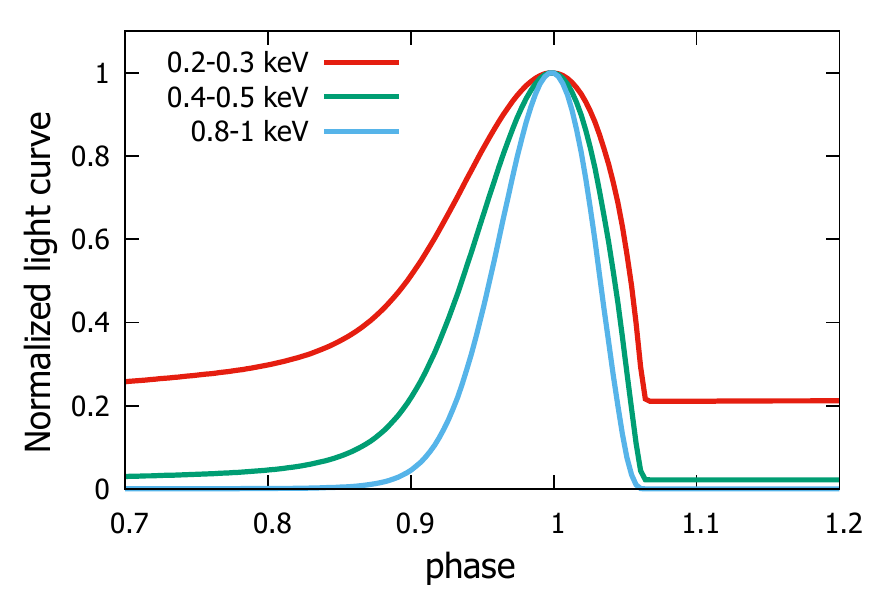}
   \caption{The light curves normalized by the peak flux in three different energy bands. The red, green and blue lines correspond to the 0.2-0.3 $\rm keV$, 0.4-0.5 $\rm keV$ and 0.8-1 $\rm keV$ light curves, respectively.}
   \label{fig:peak_profile}
\end{figure}

GSN 069 is observed to display QPEs about every 9 hours. During the outbursts, the X-ray flux can increase by about two order of magnitudes \citep{2019Natur.573..381M}. We compare the results of our model with the observed quasi-periodic 0.4-2 keV light curve of GSN 069 in Figure \ref{fig:light_curve}. It is found that our theoretical results can qualitatively fit both the period and the outburst duration, although the alternating strong/weak peaks and short/long recurrence times are not well reproduced. Furthermore, in Figure \ref{fig:peak_profile}, we present the normalized light curves for the three different energy bands given by Figure 2b of \citet{2019Natur.573..381M}. The narrower width found in high energy bands is also qualitatively consistent with observational results of GSN 069. On the other hand, QPEs in higher energy bands peak earlier in GSN 069, and our model does not reproduce this behavior.

\begin{figure}[htbp]
   \centering
   \includegraphics[width=0.47\textwidth]{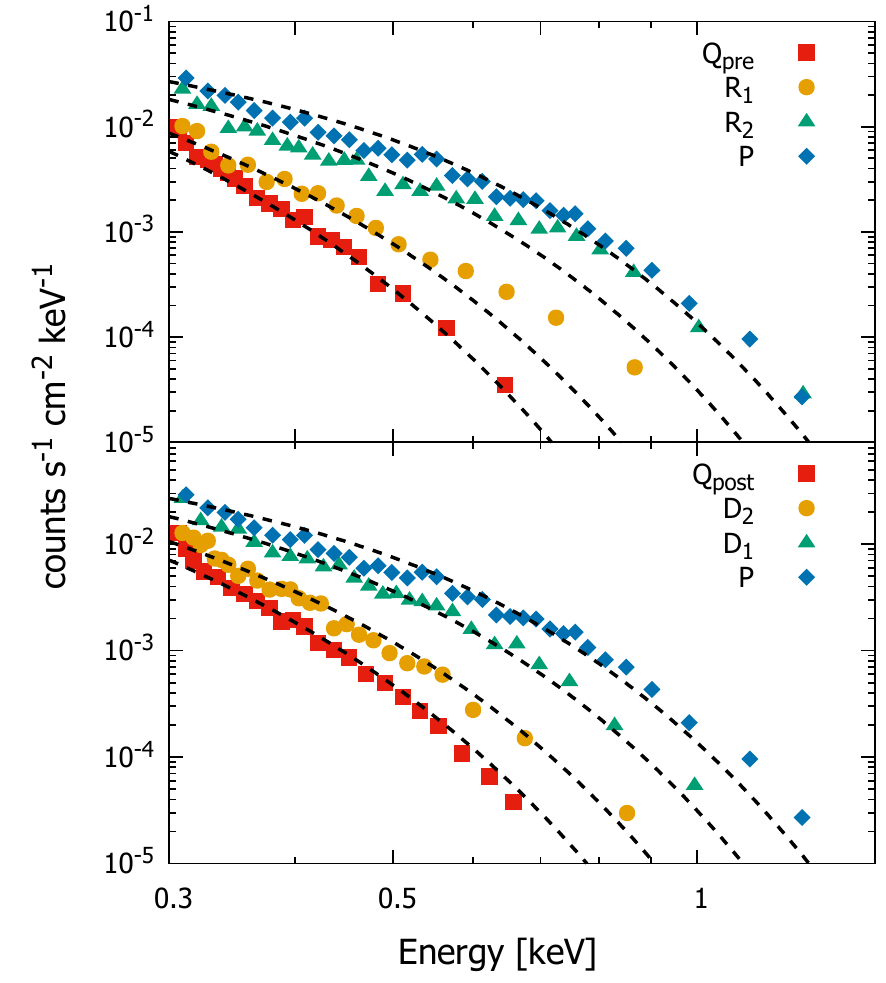}
   \caption{Phase-resolved Spectral analysis of the GSN 069. The colored points with various shapes are the observational data after doing corrections of instrumental effective area and absorption. }
   \label{fig:spectrum}
\end{figure}

\begin{figure*}[htbp]
   \centering
   \includegraphics[width=1\textwidth]{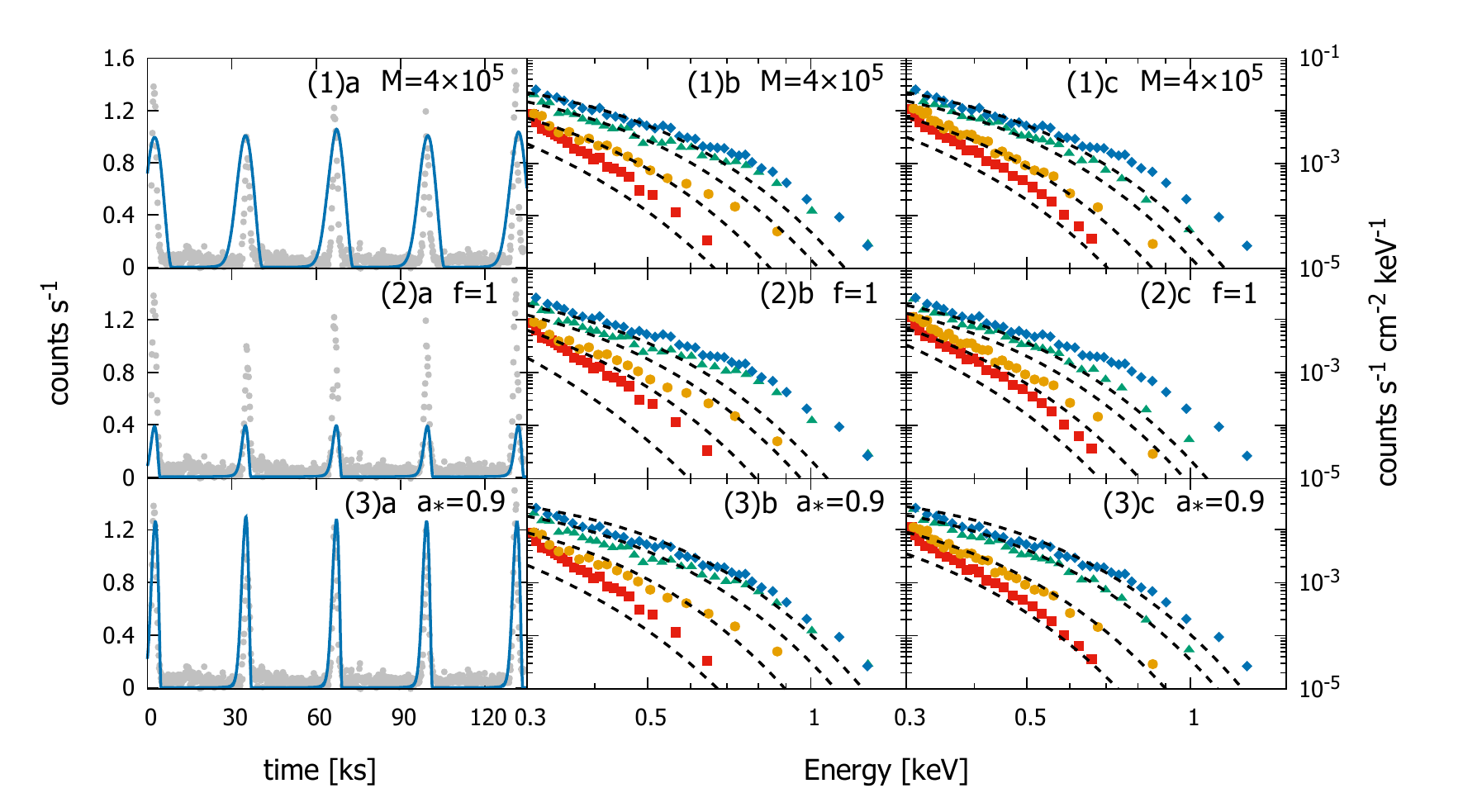}
   \caption{The 0.4-2 keV light curves and phase-resolved Spectral analysis under different parameter groups. 
   Figure \ref{fig:compare}(1), \ref{fig:compare}(2), and \ref{fig:compare}(3) correspond to $M=4\times10^5 M_{\odot}$, $f=1$(zero torque condition), and $a_*=0.9$, respectively. The best-fitting parameters of Figure \ref{fig:compare}(1), \ref{fig:compare}(2), and \ref{fig:compare}(3) correspond to Group 2, Group 3, and Group 4 in Table 1, respectively.}
   \label{fig:compare}
\end{figure*}

The timing properties of GSN 069 has been reproduced by several works (e.g., \citealt{2020MNRAS.493L.120K,2021ApJ...909...82R,2021ApJ...921L..32X}). However, besides the timing properties, the X-ray spectrum during outburst have also been well resolved by XMM-Newton (see Figure 4 in \citealt{2019Natur.573..381M}, where a whole outburst is divided into seven phases). We adopt seven phase segments of outburst ($\rm{Q_{pre}}$, $\rm{R_1}$, $\rm{R_2}$, $\rm{P}$, $\rm{D_1}$, $\rm{D_2}$, $\rm{Q_{post}}$, see the colored lines in Figure \ref{fig:light_curve}) to fit the observed X-ray spectrum. It is found that all the X-ray spectrum during outburst can be qualitatively reproduced by our model. The average effective temperature of unstable zone corresponding to the seven segments are $kT_{\rm Q_{pre}}=57\rm eV$, $kT_{\rm R_1}=67\rm eV$, $kT_{\rm R_2}=81\rm eV$, $kT_{\rm P}=91\rm eV$, $kT_{\rm D_1}=82\rm eV$, $kT_{\rm D_2}=68\rm eV$ and  $kT_{\rm Q_{post}}=59\rm eV$, respectively. The phase-resolved X-ray data in Figure \ref{fig:spectrum} are taken from Figure 3 in \citet{2019Natur.573..381M} after correcting the instrumental effective area and galaxy absorption, in which the cross-sections given by \citet{1983ApJ...270..119M} and the equivalent hydrogen column $n_{\rm H}$ ($=3 \times 10^{20} \rm cm^{-2})$ are adopted. Some of the spectra show a high-energy excess with respect to our model prediction, i.e. the agreement is qualitative but there is still room for improvement, as discussed in the following section.


\section{Choice of parameters in our model}

In addition to the width of unstable region $\Delta R$, our model include 7 free parameters, i.e., the black hole mass $M$, black hole spin $a_{*}$, $f$, mass accretion rate $\dot{m}$, $\alpha$, $\beta_{1}$ and $\mu$. $\Delta R$ is self-consistently given by the disk instability criteria given in \citet{1998Black}, which indicates that the disk is unstable when $4-10\beta_{2}-7\mu+7\beta_{2} \mu > 0$ , where $\beta_{2}$ is determined by the mass accretion rate and radius. There are five main characteristics required to fit the data points in Figures 2 and 4 well, i.e., 1) the eruption period; 2) the eruption duration and shape; 3) the amplitude of eruption; 4) the constant temperature component during spectral evolution (representing the constant outer disk emission); 5) a variable temperature component during spectral evolution. As mentioned, our goal here is to provide a possible framework for QPE production, obtaining results that are close enough, both in timing and spectral properties, to the observed data to suggest that this type of modeling is worth exploring further in the future.

The maximal disk temperature inferred from observation is $kT_{\rm max}\sim 50\rm eV$. However, for a thin accretion disk surrounding a black hole with mass of $4\times10^5 M_{\odot}$, the maximal disk temperature is only $kT_{\rm max}\sim 30\rm eV$ ($kT_{\rm{max}}=11.5\left(M/10^8M_{\odot}\right)^{-1/4}\dot{m}^{1/4} (\rm eV)$). In order to match the disk temperature before outburst (the constant temperature component), a small black hole mass $2\times10^5 M_{\odot}$, a high black hole spin ($a=0.98$) and a non-zero viscous torque are all simultaneously necessary at the first step. We notice that a spectral hardening factor $\kappa$ ($\sim1.7$) between the effective temperature of disk $T_{\rm eff}$ and observed disk temperature $T_{\rm max}$ ($T_{\rm max} \sim \kappa T_{\rm eff}$) is adopted in \citet{2019Natur.573..381M} when fitting the black hole mass. With the parameter $\kappa$, a disk temperature $T_{\rm eff}\sim 30\rm eV$ is enough to fit the observed spectra. In this case, a black hole with mass of $4\times10^5 M_{\odot}$ should be able to fit the observational data well. Our model ignores any spectral hardening, and therefore the matching with the observed constant temperature can only be obtained with a smaller black hole mass. In Figure \ref{fig:compare}(1), we investigate the effects of black hole mass $M$ on the 0.4-2 kev light curves and phased resolved spectral analysis. During the process of fitting, the eruption period and duration are the first two characteristics should be matched. It is found that the parameters with $M=4\times10^5 M_{\odot}$ (the detailed parameters adopted are listed in Table 1.) can reproduce a similar evolution trend for both the light curves and the spectral analysis, but provide a much worse match to the data. The worse fit of the spectral evolution (see Figure \ref{fig:compare}(1)b and (1)c) is driven by a too low constant temperature of the thermal component that is present in all fits, and could likely be cured by introduced a hardening factor, as explained above. On the other hand, QPEs have the right recurrence time but are too long (broad), as shown in Figure \ref{fig:compare}(1)a. However, if the mismatch of the spectral evolution fits can be ignored due to the hardening factor, the QPEs may be made narrower (shorter) by playing with a small value of $\mu$ (see Figure \ref{fig:duty_cycle}). Next, we adjust the other four parameters ($\alpha$, $\beta_1$, $\mu$ and $\dot{m}$) to reproduce the other four characteristics observed. The parameters $\mu$ can change the ratio of eruption duration to the period (see Figure \ref{fig:duty_cycle}). How the last three parameters affect the outburst behaviour had been reported by PLC21. It is found that the stronger magnetic field strength (smaller $\beta_{1}$) will result on both the decreasing eruption period and higher eruption amplitude. However, increasing $\alpha$ can only increase the eruption period (see PLC21 for details). The increase of mass accretion rate will improve the maximal disk temperature ($T_{\rm max}\sim \dot{m}^{1/4}$) and the width of unstable zone, resulting on the enhancement of all the five characteristics.

The non-zero torque in equation \ref{non_zero_torque} will emerge when considering the magnetic torque exerted on the gas at ISCO by the in-falling gas inside the ISCO \citep{1999ApJ...515L..73K}. However, this torque should be small due to the small scale-height and the weak magnetic field in a thin disk \citep{2021NewA...8501493N}, which is also suggested by GRMHD simulations \citep{2010ApJ...711..959N}. In our model, a small non-zero torque $f=0.9$ ($\sim 10\%$ of the angular momentum at ISCO) is adopted, which is consistent with the results of simulations ($f=1$ corresponding to the zero torque condition). We give the results with zero torque in Figure \ref{fig:compare}(2). It is found that the results with non-zero torque are much better than that with zero torque because the effective gas temperature close to ISCO can be greatly increased with non-zero torque (the effective temperature at ISCO is zero for zero torque).

The spin of black hole is adopted as 0.98 in this work. As discussed by \citet{1974ApJ...191..507T}, the maximal black hole spin can reach 0.998. However, the maximal value of spin will be smaller than 0.998 when taking the non-zero torque into account, which seems to be inconsistent with our assumption. However, the torque at ISCO is small as suggested by the MHD simulations and thus it will not affect the maximal value of spin significantly. Furthermore, QPEs are likely transient phenomena in the accretion history of a given source (no QPEs are observed in GSN 069 after June 2020), and if they correspond to non-zero torque transient phases, the black hole spin could well have been established during past long-lived zero-torque phases. Our model only allows a slightly lower BH spin in order to fit the observational data well (see Figure \ref{fig:compare}(3), the fitting results with $a_*=0.9$ are a little worse but still acceptable). Finally, although the width of the unstable region is small ($\Delta R=0.12 R_{\rm s}$), it can still significantly change the 0.4-2 kev light curve (Figure \ref{fig:light_curve}). However, the variation of bolometric luminosity is less than $20\%$ for the reason that the light curves are almost constant at optical-UV bands, which is consistent with observational results \citep{2019Natur.573..381M}.



\begin{table}[htbp]
    \centering
    \caption{Detailed parameter of our fitting}
    \begin{tabular*}{0.47\textwidth}{cccccccc}
    \hline
    \hline
        Number & $M$ & $a_{*}$ & $f$ & $\dot{m}$ & $\alpha$ & $\beta_{1}$ & $\mu$ \\
    \hline
        1 & $2\times10^{5}$ & 0.98 & 0.9 & 0.08 & 0.15 & 38 & 0.27 \\
        2 & $4\times10^{5}$ & 0.98 & 0.9 & 0.08 & 0.15 & 22 & 0.3 \\
        3 & $2\times10^{5}$ & 0.98 & 1 & 0.08 & 0.15 & 50 & 0.24 \\
        4 & $2\times10^{5}$ & 0.9 & 0.9 & 0.08 & 0.15 & 35 & 0.25 \\
    \hline
    \end{tabular*}
    \label{tab:pars}
\end{table}

\section{Conclusions and Discussion}

\begin{figure}[htbp]
   \centering
   \includegraphics[width=0.47\textwidth]{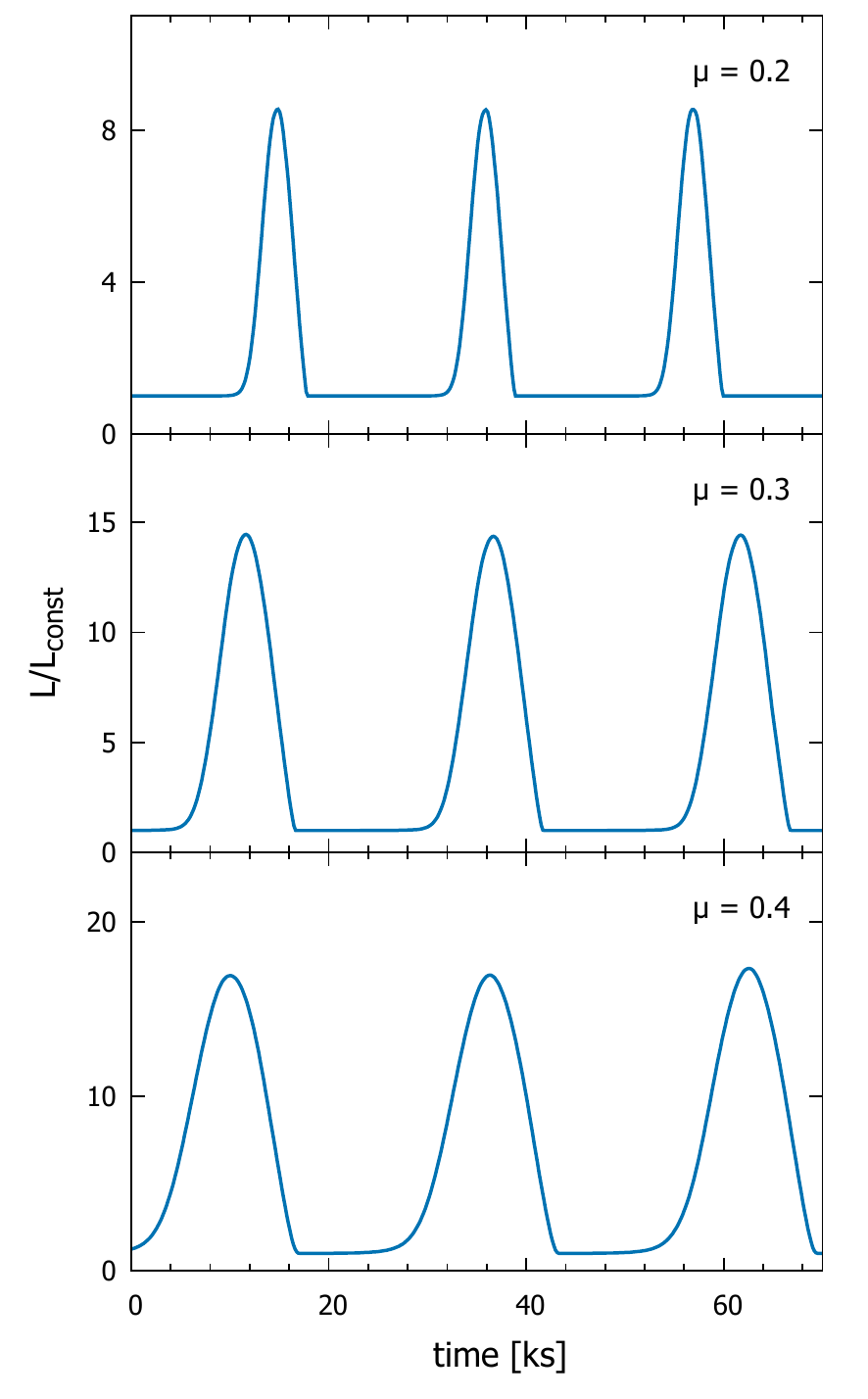}
   \caption{The 0.4-2 keV light curve, where $L_{\rm const}$ is the luminosity of outer constant disk and the value of $\mu$ in the top, middle and bottom panels are $0.2$, $0.3$ and $0.4$, respectively. Other parameters adopted in this figure are: $f=0.9$, $\dot{m}=0.1\dot{M}_{\rm Edd}$, $\alpha=0.15$, $\beta_{1}=50$ and $\Delta R=0.1 R_{\rm s}$.}
   \label{fig:duty_cycle}
\end{figure}

The mechanism of QPEs is still unclear. In this work, we adopt a disk instability model to explore the physical origin of QPEs in GSN 069. The model of PLC21 is improved in that: 1) importing the non-zero torque on the inner boundary of thin disk; 2) adopting a general form of viscous stress tensor ($T_{r\phi}=-2\alpha P_{\rm tot}^{1-\mu}P_{\rm gas}^{\mu}H$); 3) adopting a Kerr metric. Our model can qualitatively fit the 0.4-2keV light curves, the light curves at different energy bands and the phase-resolved X-ray spectrum (Figures \ref{fig:light_curve}, \ref{fig:peak_profile}, \ref{fig:spectrum}). However, we emphasize that the parameters adopted above may not represent the real best-fitting set. Exploring the full parameter space is, however, beyond the scope of this work as a full set of simulations would require several decades of computing time (by personal computer). A more quantitative comparison with data could be possible by fixing some/several free parameters to fiducial values, and by considering only a limited set of variable ones. We defer this study to future work.

The period, outburst duration and light-curve profile of QPEs are  source-dependent (see e.g., \citealt{2019Natur.573..381M,2020A&A...636L...2G,Arcodia_2021}). The period and amplitude of outburst can be changed significantly by the presence of large-scale magnetic field (PLC21). Furthermore, we suggest that the $\mu$ parameter in the general form of viscous torque can be an useful tool to fit the light-curve profile of QPEs (Figure \ref{fig:duty_cycle}). It is found that the ratio of outburst duration to the period increases with increasing $\mu$, which means that our model may also be applicable to other QPEs. Although our model can qualitatively fit the characteristics of light curves in GSN 069, some observational details are still problematic for our toy model, e.g., the asymmetric eruptions (a faster rise and slower decay is seen in some of the QPE sources); the peak delay at different energy band (the flux of higher energy bands peak at earlier times); the hard X-ray excess ($\sim 1\rm keV$) that is not well reproduced by our model during the $\rm{R_{1}}$, $\rm{R_2}$ and $\rm{P}$ phases; and the alternating long/short and strong/weak QPEs. The first two issues may be resolved if we allow the unstable zone to propagate outwards instead of remaining constant \citep{2007ApJ...666..368L}, the radiative area of unstable zone will increase after eruption and result on the the asymmetric eruptions and the peak delay at different energy band. The hard X-ray excess at $\rm{R_{1}}$, $\rm{R_2}$ and $\rm{P}$ phases may be related to the appearance of a warm comptonizing corona, as generally observed in AGN \citep{2020A&A...634A..85P}. The alternating strong/weak QPEs and long/short recurrence times may be partly solved if the unstable zone is variable instead of a constant due to the slight variation of magnetic field, perhpas responding to the QPEs themselves in a sort of feedback loop. We can expect a longer recurring time when the eruptions are stronger, which will drag more gas from accretion disk into the black hole. We will further investigate these issues in the future work.

\section* {ACKNOWLEDGEMENTS}
We thank the referee for his/her helpful suggestions and comments. This work is supported by the NSFC (grants 11773056, 11773050, 11833007, 11873073, and 12073023) and partly funded by Project No. MDM-2017-0737 Unidad de Excelencia "Mar\'{i}a de Maeztu" - Centro de Astrobiolog\'{i}a (CSIC-INTA) by MCIN/AEI/10.13039/501100011033. Support for this work was also provided by the science research grants from the China Manned Space Project with NO. CMS-CSST-2021-A06 and Shanghai Pilot Program for Basic Research - Chinese Academy of Science, Shanghai Branch (JCYJ-SHFY-2021-013).

\bibliographystyle{aasjournal}
\bibliography{ref}

\end{document}